\documentclass[letterpaper, 10 pt, conference]{ieeeconf}
\IEEEoverridecommandlockouts
\usepackage{graphicx}
\usepackage{hyperref}
\hypersetup{
    colorlinks=true,
    linkcolor=cyan,
    filecolor=magenta,
    urlcolor=cyan,
}
\usepackage{xcolor}

\title{\LARGE \bf Robots Racialized in the Likeness of Marginalized Social Identities are Subject to Greater Dehumanization than those racialized as White}

\author{Megan Strait$^{*}$, Ana S\'{a}nchez Ramos, Virginia Contreras, and Noemi Garcia$^{1}$%
\thanks{$^{1}$Department of Computer Science, The University of Texas Rio Grande Valley, Edinburg, TX USA; $^{*}$\href{megan.strait@utrgv.edu}{megan.strait@utrgv.edu}}%
}

\begin{document}

\maketitle
\thispagestyle{empty}
\pagestyle{empty}

\begin{abstract}
The emergence and spread of humanlike robots into increasingly public domains has revealed a concerning phenomenon: people's unabashed dehumanization of robots, particularly those gendered as female.
Here we examined this phenomenon further towards understanding whether other socially marginalized cues (racialization in the likeness of Asian and Black identities), like female-gendering, are associated with the manifestation of dehumanization (e.g., objectification, stereotyping) in human-robot interactions.
To that end, we analyzed free-form comments ($N=535$) on three videos, each depicting a gynoid -- Bina48, Nadine, or Yangyang -- racialized as Black, White, and Asian respectively.
As a preliminary control, we additionally analyzed commentary ($N=674$) on three videos depicting women embodying similar identity cues.
The analyses indicate that people more frequently dehumanize robots racialized as Asian and Black, than they do of robots racialized as White.
Additional, preliminary evaluation of how people's responding towards the gynoids compares to that towards other people suggests that the gynoids' ontology (as robots) further facilitates the dehumanization.
\end{abstract}

\section{INTRODUCTION}
\begin{figure}[tb!]
	\includegraphics[width=\columnwidth]{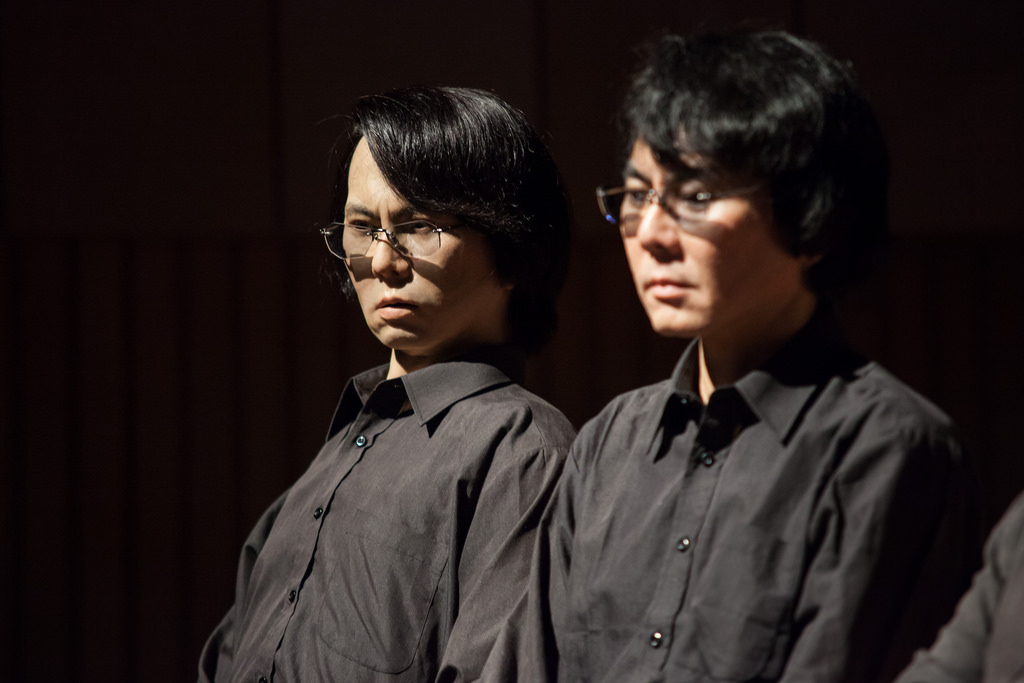}
	\caption{The Geminoid HI robot (left) alongside its originator, Hiroshi Ishiguro (right). Attribution: \href{https://flic.kr/p/fLtXv3}{photograph} available under the Creative Commons \href{https://creativecommons.org/licenses/by-nc-nd/2.0/}{Attribution-NonCommercial-NoDerivs 2.0 Generic} license.}
	\label{FIG:Geminoid}
\end{figure}

Starting in the early 2000s, a new category of robotic platforms -- androids (\cite{Ishiguro2007,MinatoEtAl2004}) -- began to emerge in academic discourse.
Characterized by their highly humanlike appearance (see, for example, Figure~\ref{FIG:Geminoid}), androids offer particular value in the degree to which they can embody social cues and effect more naturalistic interactions (e.g., \cite{HaringEtAl2016}).
In addition to the possibility of literal embodiment when used as a telepresence platform (e.g., \cite{NishioEtAL2007}), their degree of human similarity affords more realistic behaviorisms (e.g., \cite{HashimotoEtAl2007, Matsui2005}), expressivity (e.g., \cite{LeeEtAl2008}), physicality (e.g., \cite{YamashitaEtAl2017}), and overall presence (e.g., \cite{HaringEtAl2013}) than do mechanomorphic platforms.
Androids represent such a design advancement that, at first glance, they frequently ``pass'' as human (e.g., \cite{HaringEtAl2013, RosenthalVonDerPuttenEtAl2011}).

While development of androids is still in its infancy, their increasing presence in human-robot interaction (HRI) research has both underscored and enabled research on corresponding emergent human behaviors.
For example, the uncanny valley \cite{Mori1970} -- a phenomenon first noted more than 40 years ago -- only began to receive attention following the release of android platforms (e.g., \cite{MacDorman2005}).
The uncanny valley is now recognized to significantly impact HRI social outcomes \cite{KatsyriEtAl2015}, such as undermining people's trust in the agent \cite{MathurAndReichling2016} and prompting overt avoidance (e.g., \cite{StraitEtAl2015b, StraitEtAl2017b}).

More recently, in sampling the general public's perceptions of androids, we have encountered even greater cause for concern, namely: people's seeming propensity for aggression towards androids \cite{StraitEtAl2017}.
Via an evaluation of 2000 free-form comments towards a set of 24 robots (12 mechanomorphic and 12 anthropomorphic), we found aggressive tendencies to overshadow other manifestations of antisocial responding, such as the valley effect and concerns regarding a ``technology takeover''.
The prevalence of abusive responding was further exacerbated by the androids' gendering, with upwards of 40\% of commentary towards gynoids being abusive in content (evocative of gendered stereotypes, objectifying via sexualization, and/or threatening of physical harm).

\subsection{Associations between Gendering and Dehumanization}
Gender plays a powerful role in how people perceive, evaluate, and respond to others in human social interactions (e.g., \cite{Allport1954, Crenshaw1991, ItoEtAl2003}).
Even when robots lack explicit gendering, the automaticity at which people categorize and make inferences on the basis of gender nevertheless influences the human-robot interaction dynamics.
For example: gender-stereotypic cues in a robot's morphology and head-style are enough to prompt the attribution of gender to an otherwise agendered robot \cite{BernotatEtAl2017, EysselAndHegel2012}; the perception of a robot as gendered prompts different evaluations of its likability \cite{StraitEtAl2015a}; and nonconformity of a robot's behavior relative to extant stereotypes associated with its gendering reduces user acceptance \cite{TayEtAl2014}.
Thus, it is not surprising that antisocial behavior in the form of gender-based stereotyping, bias, and aggression extends to human-\emph{like} interactions with gynoids.

\begin{figure*}[tb!]
	\includegraphics[width=.33\textwidth]{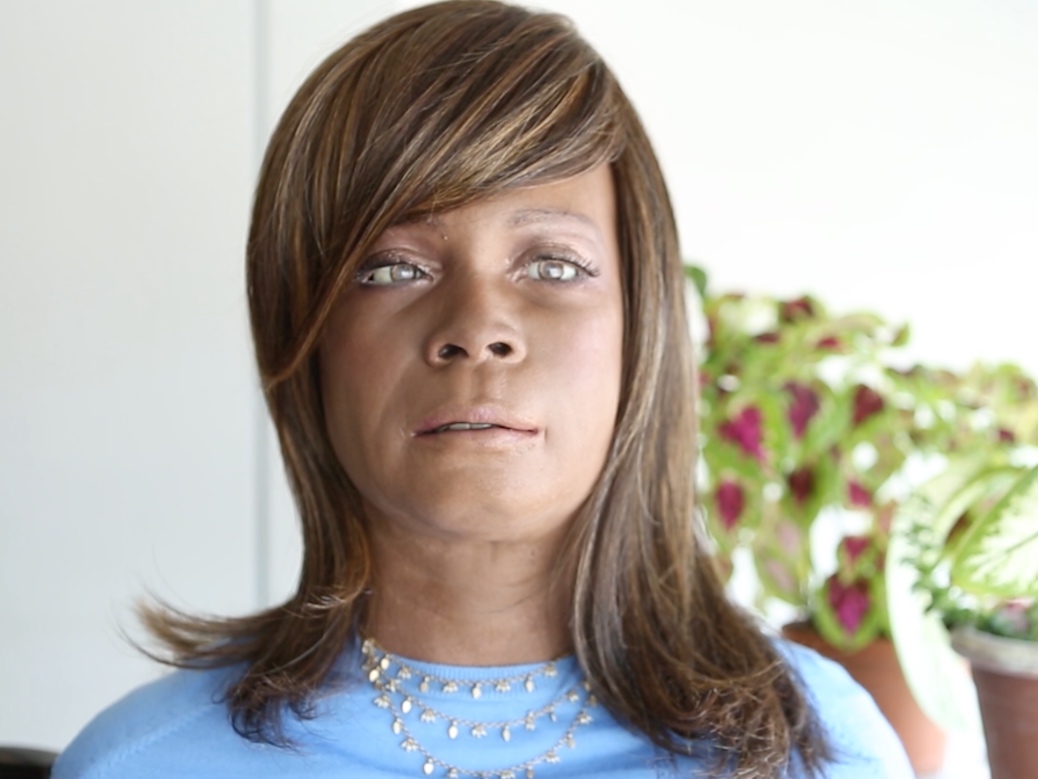}
	\includegraphics[width=.33\textwidth]{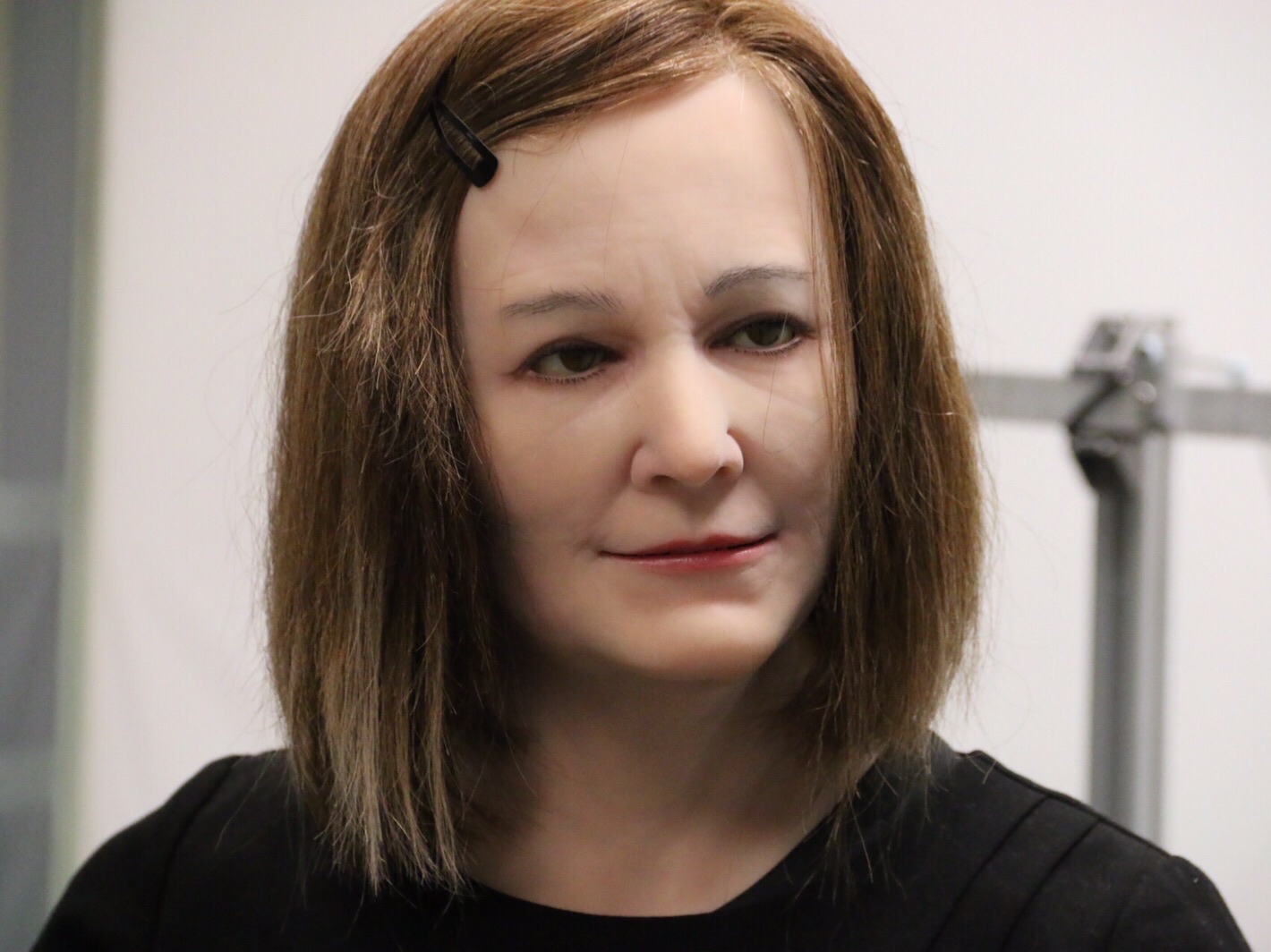}
	\includegraphics[width=.33\textwidth]{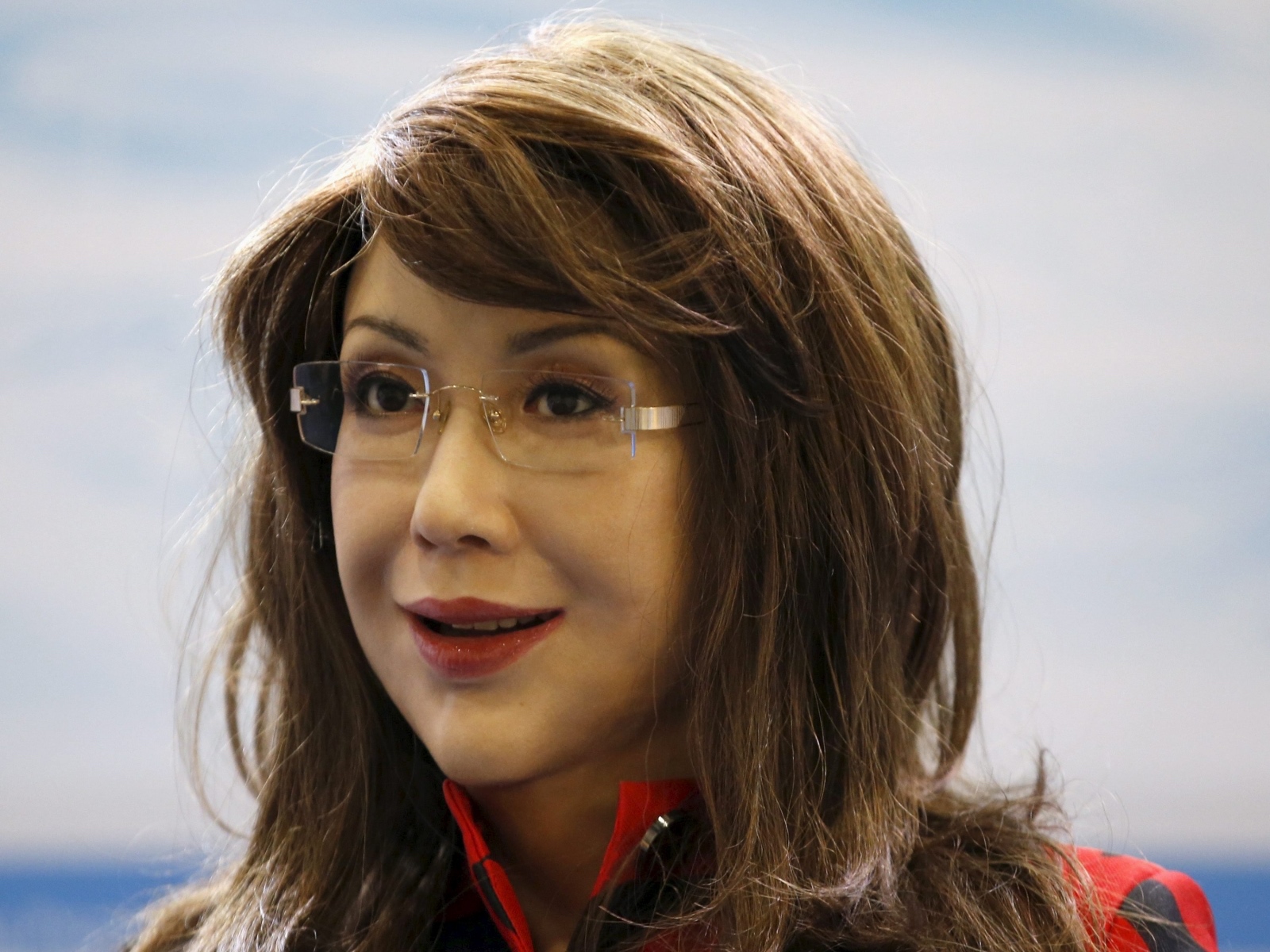}
	\caption{The three gynoids -- Bina48, Nadine, and Yangyang -- which are of varying racialization (Black, White, and Asian, respectively).}
	\label{FIG:Exemplars}
\end{figure*}

What is surprising, however, is the frequency at and degree to which aggression towards female-gendered robots manifests.
For example, in a study of a female-gendered virtual agent deployed as an educational assistant in a supervised classroom setting, more than 38\% of students were abusive in their interactions \cite{VeletsianosEtAl2008}.
In addition to general vulgarities, the students -- who were \emph{adolescents} -- often employed hypersexualizing and objectifying commentary (e.g., ``shut up u hore'', ``want to give me a blow job'', ``are you a lesbian?'').
Even in interactions with agendered agents, researchers have observed an association between the attribution of female gendering and abuse, wherein interlocutors appear to facilitate their aggression by invoking inexplicable gendering \cite{BrahnamAndDeAngeli2012}.
The observations suggest that cues of marginalizing power in human-human social dynamics can facilitate extreme dehumanization\footnote{Per strict definitions, artificial agents -- in lacking actual humanness -- cannot be dehumanized. However, as people ``humanize'' agents automatically and without intent (e.g., \cite{ReevesAndNass1996}), it is thus possible to dehumanize them.} of robots as well.
Thus, as robot designs have advanced to the point of near-human human similarity -- wherein marginalizing cues may be explicitely encoded in the agent's appearance -- there is a critical need for attention.

\subsection{Associations between Racialization and Dehumanization}
As with features associated with gender, features associated with human racial categories are encoded in the human-based design of androids.
Like gender, race (categorization based on characteristics perceived to be indicative of a particular ancestry) profoundly impacts human social dynamics.
Categorization on the basis of race, as with gender, occur automatically \cite{ItoEtAl2003} -- influencing people's attitudes and behaviors \cite{Allport1954}, often without people's awareness (\cite{Bargh1994, BarghAndChartrand1999}).

Like female-gendering, preliminary research indicates that racial cues, which are marginalizing in human-human interactions, are similarly marginalizing in human-agent interaction (e.g., \cite{BartneckEtAl2018, EysselAndHegel2012, SanchezEtAl2018}).
For example, a recent HRI study suggests that people's anti-Black/Brown behavioral biases extend to robots racialized as Black/Brown \cite{BartneckEtAl2018}.
Studies of social dynamics in interactive virtual environments (IVEs) further indicates the extensibility of racial bias (e.g., \cite{DotschAndWigboldus2008, GroomEtAl2009}).
For example, racist aversions are mirrored in people's interactions with avatars of corresponding racialization via reduced proxemics and compliance with requests (\cite{MccallEtAl2009, EastwickAndGardner2009}).

\begin{table*}[tb!]
\caption{Source information of the six videos from which public commentary was scraped.}
\label{TABLE:VIDEOS}
\centering
\begin{tabular}{ccc cccccc}
\hline\noalign{\smallskip}
\textbf{Agent} 			& \textbf{Racialization}
                        & \textbf{Source Video} & \textbf{Release} & \textbf{Duration}
						& \textbf{Views}  		& \textbf{Likes}   & \textbf{Dislikes}
						& \textbf{Comments}\\
\noalign{\smallskip}
\hline
\\
\textbf{Bina48} 		& Black & \url{youtu.be/G9uLnquaC84} & $2015$ & $00:01:58$
						& $201K$ 	& $251$ & $67$ & $123$\\
						%& $90$\\

\textbf{Nadine} 		& White & \url{youtu.be/cvbJGZf-raY} & $2015$ & $00:01:10$
						& $331K$ 	& $411$ & $63$ & $250$\\
						%& $162$\\

\textbf{YangYang} 		& Asian & \url{youtu.be/K53t27U1FC0} & $2015$ & $00:03:56$
						& $319K$ 	& $374$ & $80$ & $162$\\
						%& $76$\\
\\
\hline
\\
\textbf{Beyonce	Knowles}& Black & \url{youtu.be/mOHYTfVXGVc} & $2013$ & $00:03:52$
						& $763K$ 	& $3.4K$ & $111$ & $383$\\
						%& $181$\\
\textbf{Cameron Diaz}	& White & \url{youtu.be/e-HvL3TSf-8} & $2015$ & $00:05:29$
						& $681K$ 	& $4.7K$ & $164$ & $222$\\
						%& $122$\\
\textbf{Liu Wen}		& Asian & \url{youtu.be/dfLZV00r8W4} & $2015$ & $00:04:51$
						& $156K$ 	& $1.3K$ & $16$ & $69$\\
						%& $46$\\
\\
\hline
\end{tabular}
\end{table*}

\subsection{Present Work}
Given the frequency and degree to which robot-direct abuse has been observed, it is important to understand the associations between antisociality and the identity cues embodied by the agent.
Specifically, knowing when, how, and why people respond to gendering and racialization can facilitate the interpretation and mitigation of aggression in human-robot interactions.
The present lack otherwise stands in the way of meaningful and appropriate social interactions with robots.
Furthermore, left unaddressed, aggression in these contexts has the potential to reinforces harmful biases in human-human interactions (e.g., \cite{Sabanovic2010}).
We thus pursued an extension of our prior work investigating people's frequent dehumanization of gynoids (\cite{StraitEtAl2017}) to consider the compounding impact of race-stereotypic cues.

Via an observational study of people's responding towards robots of varying racialization, we evaluated whether racial biases and overt racism extend to people's interactions with robots racialized in the likeness of marginalized social identities.
To that end, we sampled public commentary on three online videos -- depicting Bina48, Nadine, and Yangyang -- available via \href{youtube.com}{YouTube}.\footnote{Given limited accessibility of racialized robots for and impractical methodological overhead required in a multi-platform, in-person HRI experiment, the online, observational methods were necessary to pursue the given research. Although the approach may possess less ecological validity than one involving direct, in-person human-robot interactions, comparisons of results obtained from in-person versus online testing show little difference (see, for example, \cite{CaslerEtAl2013}). This approach, in addition to enabling experimentation that is otherwise infeasible, offered further benefit via broader sampling.}
Based on human social psychology literature (e.g., \cite{Allport1954}), we expected that people would exhibit more frequent and extreme dehumanization of robots racialized as Asian and Black relative to robots racialized as White.

Finding support for this hypothesis, we then conducted a preliminary assessment as to whether people's responding towards the gynoids is a reflection of the sampling context (online social fora, which are marked by general verbal disinhibition \cite{Suler2004} and hostility towards women \cite{FoxAndTang2014, Herring1996, HudsonBruckman2002}) or whether there is any facilitation by the gynoids' non-human ontological categorization.
Specifically, we sought to test how similar people's responding to the three gynoids was in relation to women of similar identity characteristics.
To that end, we additionally sampled commentary on three videos depicting women of comparable ages and racial identities to those embodied by the three gynoids.

In total, we investigated the following research questions:
\emph{Do people more readily engage in the dehumanization of robots racialized in the likeness of marginalized social identities?} (\textbf{RQ1}); and \emph{Does people's dehumanization of robots differ from that of other humans?} (\textbf{RQ2}).

\section{Method}
\subsection{Independent Variables}
We effected a quasi-manipulation of robot \textbf{racialization} (three levels: \emph{Asian}, \emph{Black}, \emph{White}) via selection of videos from those of existing androids.
As androids are modeled after actual people, their designs encode cues associated with racial identity.
In turn, people attribute (human) racial identity to such robots (see, for example, \hyperref[FIG:Geminoid]{Figure~\ref{FIG:Geminoid}} which depicts Hiroshi Ishiguro, who is Asian, alongside the Geminoid HI, which is racialized as Asian).
A second quasi-manipulation, \textbf{ontological category} (two levels: \emph{human}, \emph{robot}), was carried out via selection of a second set of videos depicting people of corresponding racial identities.
This manipulation enabled direct, albeit preliminary, comparison of people's online behavior towards the gynoids versus other people.

\begin{figure*}[tb!]
\centering
	\includegraphics[width=.49\textwidth]{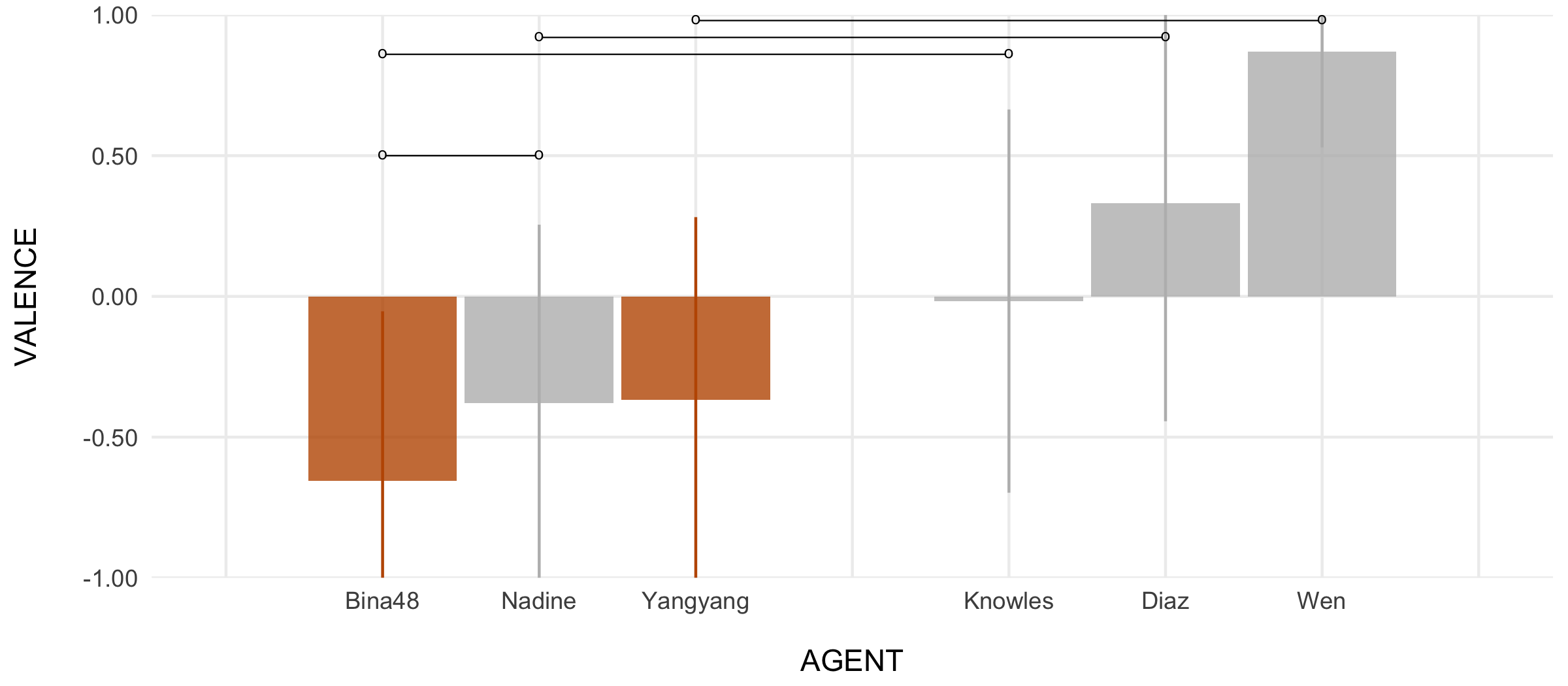}
	\includegraphics[width=.49\textwidth]{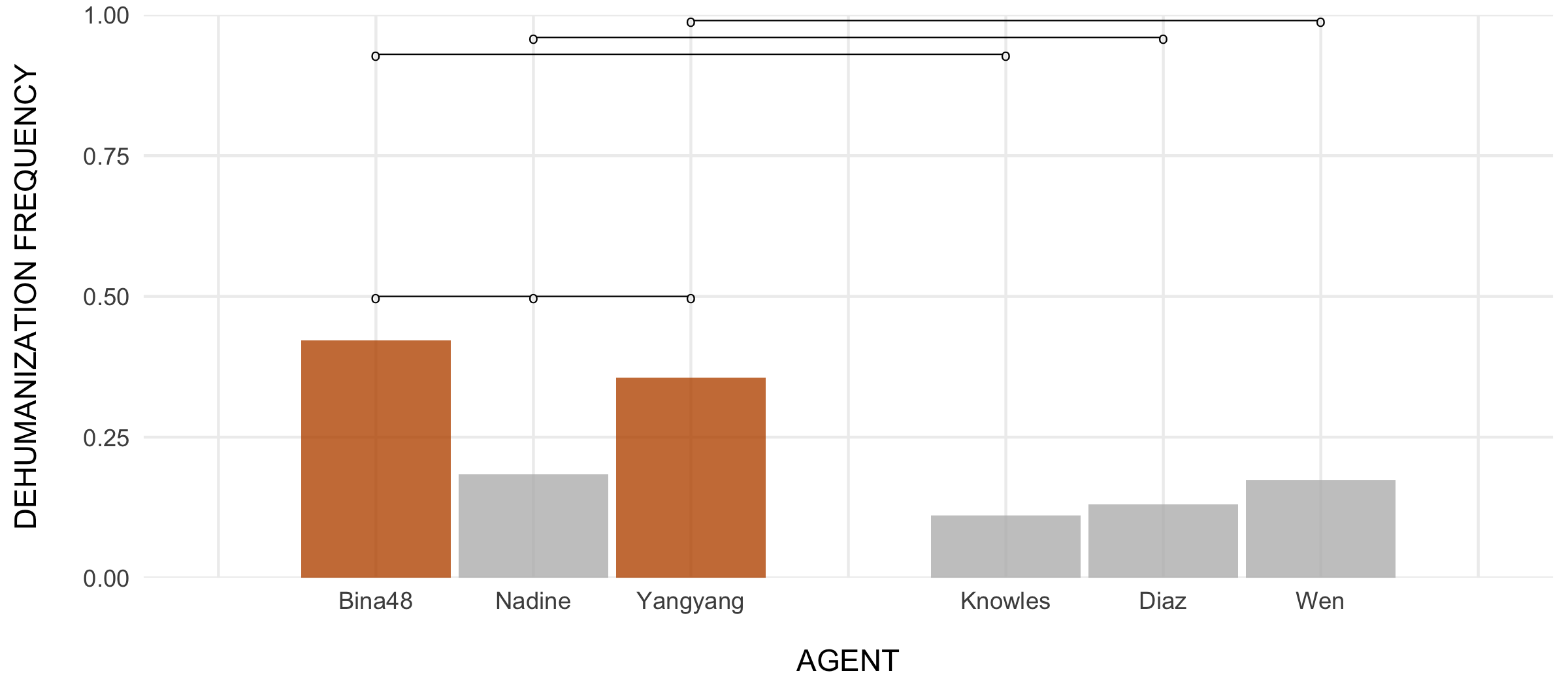}
	\caption{Mean valence (+/- SD; \emph{left}) and dehumanization frequency (\emph{right}) in response to each of the six agents (from left-to-right: Bina48, Nadine, and Yangyang; Knowles, Diaz, and Wen). Bars indicate significance (amongst the planned contrasts).}
	\label{FIG:Results}
\end{figure*}

\subsection{Materials}
A total of six videos were selected for inclusion in the present study (see \hyperref[TABLE:VIDEOS]{Table~\ref{TABLE:VIDEOS}}).
The exact selections and number of instances were driven by contraints of the current design space of humanoid robots.
Specifically, to our knowledge, four racial identities (Asian, Black, Brown, White) are represented within the set of existing androids.
There, however, exists just one platform racialized as Black (Bina48, who is female-gendered) and one racialized as Brown (Ibn Sina, who is male-gendered).
To avoid gender-based confounds due to intersecting marginalization (see \cite{Crenshaw1991}), we optimized for representation of racial identity while holding robot gendering constant (to female-presenting).
Limiting the search to gynoids, we then selected for robots most similar in approximate age (as reflected by their appearance).
This yielded three gynoids -- Bina48, Nadine, and Yangyang (see \hyperref[FIG:Exemplars]{Figure~\ref{FIG:Exemplars}}) -- for which we then identified three videos of similar content and metrics (release year, views, etc.).

Subsequently, we conducted a search to identify three videos depicting an Asian, Black, and White woman for comparison to the three gynoids.
We again attempted to control for (i.e., maximize similarity in) age, video content, and video metrics.
Due to the relative publicity of emerging robot platforms (e.g., gynoids), we focused our search towards women of moderate celebrity and ultimately selected:
an interview-style video of musician, Beyonce Knowles; a brief presentation by actor, Cameron Diaz\footnote{While Diaz is of Spanish/Cuban ancestry, she is White-presenting.} (for comparison with Nadine); and a featurette of model, Liu Wen, for comparison with Bina48, Nadine, and Yangyang respectively.

\subsection{Data Acquisition \& Retention}
All available comments from the six videos ($N=1209$) were retrieved from their respective sources on September 1, 2017.
The comments were then preprocessed as follows:
comments not in English or that were duplicates, indecipherable\footnote{For example, the comment, ``FEMALES MOST LIKELY BE ON MARS[...]'', on the video depicting Yangyang was discarded.}, or unrelated\footnote{For example, a comment from a person correcting the grammar of another comment, ``[@]Joel Marx The past tense is 'began'.''), on the video depicting Nadine was discarded.} to the video content were discarded; sequential comments written by a single user (without interruption by replies and in a single timeframe) were condensed and treated as one.
In total, $677$ comments ($N_{Bina}=90$, $N_{Nadine}=162$), $N_{Yangyang}=76$); $N_{Knowles}=181$), $N_{Diaz}=122$, $N_{Wen}=46$) were retained for analysis.

\subsection{Dependent Variables}
The $677$ retained comments were then each coded on two dimensions by six research assistants trained in coding, but blind to the hypotheses.
Specifically, comments were coded for the \textbf{valence} of the response (positive, neutral, or negative; Fleiss' $\kappa=.86$), and presence (0 or 1) of dehumanizing commentary ($\kappa=.83$), which was used to compute an overall \textbf{frequency of dehumanization}.
Commentary was coded as dehumanizing if it contained content that was objectifying (including overt sexualization, \cite{MoradiAndHuang2008}, as well as ambivalent sexism \cite{GlickAndFiske1996}), racist (i.e., evocative of race-based stereotypes \cite{Allport1954}), and/or abusive (i.e., descriptive of verbal hostility or physical violence) towards the given agent.

\section{Results}
\hyperref[FIG:Results]{Figure \ref{FIG:Results}} shows the mean valence and frequency of dehumanizing commentary for each of the six agents.
All contrasts were evaluated at a significance level of $\alpha=.05$.
Welch's t-tests\footnote{Welch's t-test is an adaptation (of Student's) for testing whether two independent populations have equal means, without assuming equal variance. It is more reliable than Student's t-test and thus used when populations have unequal variance and sample sizes \cite{Ruxton2006}, as was the case in the present study.} were used to compare the valence of people's responding across agents; the effect sizes for significant contrast are reported in terms of Cohen's $d$.
To compare the proportions of dehumanizing commentary accross agents we used exact binomial tests (due to the binary property of the measure).
Correspondingly, ``effect size'' of significant binomial tests were reported in terms of $RR$ (relative risk).

\subsection{RQ1: Effects of Racialization}
The association between marginalizing racialization and antisocial responding towards robots was evaluated via two planned, one-tailed contrasts: responding towards Nadine (racialized as White) versus responding towards Bina48 and towards Yangyang (each of which are racialized in the likeness of identities associated with social marginalization).
Specifically, based on prior research, we hypothesized that people would exhibit greater antisociality towards Bina48 and Yangyang than they would towards Nadine.

Overall, the valence of people's commentary on the three gynoids was negative irrespective of the specific robot ($M=-.47$, $SD=.16$).
In particular, there was no significant difference in the valence of people's responding between Nadine ($M=-.38$, $SD=.63$) and Yangyang ($M=-.37$, $SD=.65$; $t=.12$, $p=.89$).
However, consistent with prior findings indicating the extension of anti-Black/Brown biases to HRI (e.g., \cite{BartneckEtAl2018}), commentary was significantly more negative in response to Bina48 ($M=-.66$, $SD=.60$) than in response to Nadine ($t=3.35$, $p<.01$, $d=.44$).
Similarly, across the three gynoids, a substantial proportion of people's commentary was dehumanizing in content ($M=.32$, $SD=.12$).
However, both Bina48 ($M=.42$) and Yangyang ($M=.36$) were subject to significantly more dehumanizing commentary than was Nadine ($M=.18$; $p<.01$, $RR_{Bina48}=2.30$, $RR_{Yangyang}=1.93$).

\subsection{RQ2: Effects of Ontology}
We next evaluated whether people's responding towards the three gynoids was a reflection of the sampling context (e.g., due to online disinhibition \cite{Suler2004}) or whether the degree of antisociality is facilitated by the gynoids' non-human ontological identity (as robots).
For each of the two measures (valence, dehumanization frequency) we computed three planned, two-tailed contrasts of the robot versus human ontological categories: responding towards Bina48 versus Knowles, Nadine versus Diaz, and Yang Yang versus Wen.

Overall, the valence of commentary on the three women was positive ($M=.39$, $SD=.45$).
Contrasted with commentary on the corresponding gynoids, responding was significantly more negative ($p<.01$) towards the robots:
\begin{itemize}
\item Bina48 vs. Knowles: $t=7.55$, $d=.99$
\item Nadine vs. Diaz: $t=8.43$, $d=1.00$
\item Yangyang vs. Wen: $t=12.06$, $d=2.38$
\end{itemize}

Despite an overall positive valence, a non-negligable proportion of people's commentary on the three women was dehumanizing in content ($M=.14$, $SD=.03$).
However, the relative proportions in response to the women were markedly less ($p<.01$) than in response to the gynoids:
\begin{itemize}
\item Bina48 ($M=.42$) vs. Knowles ($M=.11$): $RR=3.82$
\item Nadine ($M=.18$) vs. Diaz ($M=.13$): $RR=1.40$
\item Yangyang ($M=.36$) vs. Wen ($M=.17$): $RR=2.04$
\end{itemize}

\section{Discussion}
Here we examined the associations between racialization, ontology, and the manifestation of antisocial responding towards emergent robot platforms.
Motivated by prior research, which highlights the automaticity at which bias and stereotyping extend to HRI (e.g., \cite{ReevesAndNass1996, BartneckEtAl2018}, we evaluated the degree to which people responded negatively and dehumanizingly towards robots of varying racialization (Asian, Black, and White) versus other people of similar social cues.

\subsection{Summary of Findings}
\textbf{Do people more readily dehumanize robots racialized in the likeness of marginalized social identities?}
Across both measures (valence and dehumanization frequency), the data show a marked difference in the degree to which people respond to a gynoid racialized as Black versus one racialized as White.
While the data in response to Yangyang (versus Nadine) is mixed, with a non-significant difference in valence but significant difference in the frequency of dehumanization, qualitative analysis of the commentary supports an interpretation of greater antisociality towards the Asian gynoid.
Specifically, with the proportion of dehumanizing commentary towards Yangyang nearly double that of Nadine, the data indicate that, like Bina48, people readily marginalize Yangyang.
The valence of the commentary, however, indicates that the manifestation thereof is less hostile than that towards Bina48, with many of the comments containing content that is dehumanizing but delievered with a valence that is neutral to positive.
For example, the comment -- \emph{``Wow that's cool! But the real question is... Can you fuck it?''} -- is positive overall (due to ``wow that's cool!'').
Nevertheless, the content is dehumanizing (``Can you fuck it?'').
Taken together, the data indicate general support for our hypothesis: that people readily extend racial biases and employ stereotypes to dehumanize robots implicitly racialized in the likeness of marginalized human identities.

\textbf{Does people's dehumanization of robots differ from that of other people?}
The comparison to people's commentary on women of similar identity cues to those of the robots suggests that such responding is not simply normative behavior for the context (i.e., online disinhibition; e.g., \cite{Suler2004, WotanisMcmillan2014}).
Specifically, across all three racializations, people were consistently and significantly more negative towards and dehumanizing of the gynoids relative to their human counterparts.
Although these findings are preliminary and subject to important limitations, if replicated and extended, they would suggest that antisocial responding is further facilitated by the robots' lack of actual human membership.

\subsection{Links to Existing Literature \& Broader Implications}
Here we observed that: (1) racial cues, which can be socially marginalizing in human-human interactions, are associated with more negative commentary towards and a higher frequency of dehumanization of robots embodying such racializations; (2) people's responding does not appear to be a mere function of the online context in which the data was gathered.
Rather, the data suggest that the agents' ontological categorization (as robots) -- despite their highly humanlike appearances -- facilitates greater dehumanization.

These findings are consistent with prior research indicating the automaticity at which social biases extend to and affect behavior in HRI (e.g., \cite{StraitEtAl2017, BernotatEtAl2017, EysselAndHegel2012, EysselAndLoughnan2013, BartneckEtAl2018}).
Moreover, the findings support indications by a growing body of literature (containing instances of unprovoked abuse towards robots -- e.g., \cite{BrscicEtAl2015, SalviniEtAl2010}; as well as less empathy for robots relative to that for people when witnessing or participating in their abuse -- e.g., \cite{BartneckEtAl2005, RosenthalEtAl2013}) that people more readily engage in the dehumanization of robots.
For example, during deployment of a service robot in an open, public environment, Salvini and colleagues observed people's interactions with the robot often escalated, without provocation, into physical abuse involving \emph{kicking, punching, and slapping} the robot \cite{SalviniEtAl2010}.
Brsci\'c and colleagues observed similarly violent behavior \emph{from children} in their 2015 deployment of the Robovie robot in a shopping mall \cite{BrscicEtAl2015}.
%While data on the degree to which people's antisociality towards robots mirrors that towards other people (e.g., \cite{ReaEtAl2015, ReichAndEyssel2017}), the manifesation of unprovoked abuse and early in development suggest that this antisociality may be rather deeply entrenched in how people perceive robots.

Considered alongside existing literature, the findings raise several considerations for the design and development of future robots.
For example, if a given robot is designed to learn from its interactions with people, guided and/or supervised learning (e.g., \cite{Breazeal2004}) is warranted to avoid outcomes such as the robot developing dehumanizing tendencies.
Moreover, if people are comfortable dehumanizing robots (as the present findings suggest) and do so while perceiving the robots as \emph{human}-like, this may shape people's subsequent interactions with other people if left unaddressed (e.g., \cite{Sabanovic2010}).

To that end, advancing the social capacities of robots to include the ability to detect/recognize such antisociality is necessary, as is understanding of what would serve as effective robot responses.
Three preliminary forrays exist:
(1) Towards mitigating the frequency of robot abuse, Brscic and colleagues' implemented an avoidance-based response mechanism that proactively reduces a robot's proximity to probable abusers \cite{BrscicEtAl2015}.
(2) Towards understanding effective reactive behaviors for responding to abuse, Tan and colleagues explored three common strategies (ignoring the abuser, explicit disengagement, and solicitation of empathy; \cite{TanEtAl2018}).
(3) Towards mediating aggression in multi-person human-robot interactions, Jung and colleagues explored trialed a robot's use of verbal responses grounded in counseling literature on mediating human-human conflicts \cite{JungEtAl2015}.
Each advance understanding of predictive factors and mechanisms for responding, however, further research is warranted.

\subsection{Limitations \& Avenues for Future Research}
While the present study provides an initial evaluation of the associations between robot racialization and dehumanizing responses, there are a number of methodological limitations necessitating further consideration.
In particular, we note the preliminary nature of RQ2 (effect of ontology).
The inclusion of R2 served to indicate whether antisocial responding in the context of \href{youtube.com}{YouTube} is due to general online disinhibition \cite{Suler2004}, or is otherwise different from how people respond to other people.
Nevertheless, the specific materials used may capture different responding than the average commentary on \href{youtube.com}{YouTube}, wherein the relative celebrity of the women depicted may promote more positive responding than people might show towards non-celebrity women.
Thus, replication of RQ2 with non-celebrity exemplars is needed.

\section{Conclusions}
The aim of the present work was to investigate the ways in which people respond to cues of gender and race when explicitly encoded in the appearance of a humanoid robot.
Consistent with prior research, we observed an association between racial cues marginalizing in human social dynamics and antisociality.
Specifically, people exhibited more negative and more frequently dehumanizing responding towards Bina48 and Yangyang, which are racialized as Black and Asian, relative to Nadine (which is racialized as White).
In addition, people appear to more readily engage in this manner towards robots (versus towards other people), suggesting that racial biases both extend to and are amplified in the realm of HRI.
However, further research is needed towards replicating these findings and fully understanding the social impacts of such antisociality.

\bibliographystyle{Template-IEEE/IEEEtran.bst}
\bibliography{bibliography}

\begin{thebibliography}{10}
\providecommand{\url}[1]{#1}
\csname url@rmstyle\endcsname
\providecommand{\newblock}{\relax}
\providecommand{\bibinfo}[2]{#2}
\providecommand\BIBentrySTDinterwordspacing{\spaceskip=0pt\relax}
\providecommand\BIBentryALTinterwordstretchfactor{4}
\providecommand\BIBentryALTinterwordspacing{\spaceskip=\fontdimen2\font plus
\BIBentryALTinterwordstretchfactor\fontdimen3\font minus
  \fontdimen4\font\relax}
\providecommand\BIBforeignlanguage[2]{{%
\expandafter\ifx\csname l@#1\endcsname\relax
\typeout{** WARNING: IEEEtran.bst: No hyphenation pattern has been}%
\typeout{** loaded for the language `#1'. Using the pattern for}%
\typeout{** the default language instead.}%
\else
\language=\csname l@#1\endcsname
\fi
#2}}

\bibitem{Ishiguro2007}
H.~Ishiguro, ``Android science,'' in \emph{Robotics Research}, 2007.

\bibitem{MinatoEtAl2004}
T.~Minato, M.~Shimada, H.~Ishiguro, and S.~Itakura, ``Development of an android
  robot for studying human-robot interaction,'' in \emph{International
  Conference on Industrial, Engineering and Other Applications of Applied
  Intelligent Systems}, 2004.

\bibitem{HaringEtAl2016}
K.~S. Haring, D.~Silvera-Tawil, T.~Takahashi, K.~Watanabe, and M.~Velonaki,
  ``How people perceive different robot types: A direct comparison of an
  android, humanoid, and non-biomimetic robot,'' in \emph{Proc. the
  International Conference on Knowledge and Smart Technology}, 2016.

\bibitem{NishioEtAL2007}
S.~Nishio, H.~Ishiguro, and N.~Hagita, ``Can a teleoperated android represent
  personal presence?—a case study with children,'' \emph{Psychologia}, 2007.

\bibitem{HashimotoEtAl2007}
T.~Hashimoto, S.~Hiramatsu, T.~Tsuji, and H.~Kobayashi, ``Realization and
  evaluation of realistic nod with receptionist robot saya,'' in \emph{Proc.
  RO-MAN}, 2007.

\bibitem{Matsui2005}
D.~Matsui, T.~Minato, K.~F. MacDorman, and H.~Ishiguro, ``Generating natural
  motion in an android by mapping human motion,'' in \emph{Proc. IROS}, 2005.

\bibitem{LeeEtAl2008}
D.-W. Lee, T.-G. Lee, B.~So, M.~Choi, E.-C. Shin, K.~Yang, M.~Back, H.-S. Kim,
  and H.-G. Lee, ``Development of an android for emotional expression and human
  interaction,'' in \emph{Proc. the International Federation of Automatic
  Control}, 2008.

\bibitem{YamashitaEtAl2017}
Y.~Yamashita, H.~Ishihara, T.~Ikeda, and M.~Asada, ``Appearance of a robot
  influences causal relationship between touch sensation and the personality
  impression,'' in \emph{Proc. HAI}, 2017.

\bibitem{HaringEtAl2013}
K.~S. Haring, Y.~Matsumoto, and K.~Watanabe, ``How do people perceive and trust
  a lifelike robot,'' in \emph{Proc. the world congress on engineering and
  computer science}, 2013.

\bibitem{RosenthalVonDerPuttenEtAl2011}
A.~M. von~der P{\"u}tten, N.~C. Kr{\"a}mer, C.~Becker-Asano, and H.~Ishiguro,
  ``An android in the field,'' in \emph{Proc. HRI}, 2011.

\bibitem{Mori1970}
M.~Mori, K.~F. MacDorman, and N.~Kageki, ``The uncanny valley [from the
  field],'' \emph{IEEE Robotics \& Automation Magazine}, 2012.

\bibitem{MacDorman2005}
K.~F. MacDorman, ``Androids as an experimental apparatus: Why is there an
  uncanny valley and can we exploit it,'' in \emph{Proc. COGSCI: Workshop on
  Social Mechanisms of Android Science}, 2005.

\bibitem{KatsyriEtAl2015}
J.~K{\"a}tsyri, K.~F{\"o}rger, M.~M{\"a}k{\"a}r{\"a}inen, and T.~Takala, ``A
  review of empirical evidence on different uncanny valley hypotheses: support
  for perceptual mismatch as one road to the valley of eeriness,''
  \emph{Frontiers in Psychology}, 2015.

\bibitem{MathurAndReichling2016}
M.~B. Mathur and D.~B. Reichling, ``Navigating a social world with robot
  partners: A quantitative cartography of the uncanny valley,''
  \emph{Cognition}, 2016.

\bibitem{StraitEtAl2015b}
M.~Strait, L.~Vujovic, V.~Floerke, M.~Scheutz, and H.~Urry, ``Too much
  humanness for human-robot interaction: exposure to highly humanlike robots
  elicits aversive responding in observers,'' in \emph{Proc. CHI}, 2015.

\bibitem{StraitEtAl2017b}
M.~K. Strait, V.~A. Floerke, W.~Ju, K.~Maddox, J.~D. Remedios, M.~F. Jung, and
  H.~L. Urry, ``Understanding the uncanny: Both atypical features and category
  ambiguity provoke aversion toward humanlike robots,'' \emph{Frontiers in
  Psychology}, 2017.

\bibitem{StraitEtAl2017}
M.~Strait, C.~Aguillon, V.~Contreras, and N.~Garcia, ``Online social commentary
  reflects an appearance-based uncanny valley, a general fear of a ``technology
  takeover'', and the unabashed sexualization of female-gendered robots,'' in
  \emph{Proc. RO-MAN}, 2017.

\bibitem{Allport1954}
G.~W. Allport, ``The nature of prejudice,'' 1954.

\bibitem{Crenshaw1991}
K.~Crenshaw, ``Mapping the margins: Intersectionality, identity politics, and
  violence against women of color,'' \emph{Stanford Law Review}, 1991.

\bibitem{ItoEtAl2003}
T.~A. Ito and G.~R. Urland, ``Race and gender on the brain: electrocortical
  measures of attention to the race and gender of multiply categorizable
  individuals.'' \emph{Journal of Personality and Social Psychology}, 2003.

\bibitem{BernotatEtAl2017}
J.~Bernotat, F.~Eyssel, and J.~Sachse, ``Shape it--the influence of robot body
  shape on gender perception in robots,'' in \emph{Proc. ICSR}, 2017.

\bibitem{EysselAndHegel2012}
F.~Eyssel and F.~Hegel, ``(s) he's got the look: gender stereotyping of
  robots,'' \emph{Journal of Applied Social Psychology}, 2012.

\bibitem{StraitEtAl2015a}
M.~Strait, P.~Briggs, and M.~Scheutz, ``Gender, more so than age, modulates
  positive perceptions of language-based human-robot interactions,'' in
  \emph{Proc. AISB}, 2015.

\bibitem{TayEtAl2014}
B.~Tay, Y.~Jung, and T.~Park, ``When stereotypes meet robots: the double-edge
  sword of robot gender and personality in human--robot interaction,''
  \emph{Computers in Human Behavior}, 2014.

\bibitem{VeletsianosEtAl2008}
G.~Veletsianos, C.~Scharber, and A.~Doering, ``When sex, drugs, and violence
  enter the classroom: Conversations between adolescents and a female
  pedagogical agent,'' \emph{Interacting with Computers}, 2008.

\bibitem{BrahnamAndDeAngeli2012}
S.~Brahnam and A.~De~Angeli, ``Gender affordances of conversational agents,''
  \emph{Interacting with Computers}, 2012.

\bibitem{ReevesAndNass1996}
B.~Reeves and C.~Nass, \emph{How people treat computers, television, and new
  media like real people and places}, 1996.

\bibitem{Bargh1994}
J.~A. Bargh, ``The four horsemen of automaticity: Awareness, intention,
  efficiency, and control in social cognition,'' \emph{Handbook of Social
  Cognition}, 1994.

\bibitem{BarghAndChartrand1999}
J.~A. Bargh and T.~L. Chartrand, ``The unbearable automaticity of being,''
  \emph{American Psychologist}, 1999.

\bibitem{BartneckEtAl2018}
C.~Bartneck, K.~Yogeeswaran, Q.~M. Ser, G.~Woodward, R.~Sparrow, S.~Wang, and
  F.~Eyssel, ``Robots and racism,'' in \emph{Proc. HRI}, 2018.

\bibitem{SanchezEtAl2018}
A.~C. S{\'a}nchez~Ramos, V.~Contreras, A.~Santos, C.~Aguillon, N.~Garcia, J.~D.
  Rodriguez, I.~Amaya~Vazquez, and M.~K. Strait, ``A preliminary study of the
  effects of racialization and humanness on the verbal abuse of female-gendered
  robots,'' in \emph{Proc. HRI}, 2018.

\bibitem{DotschAndWigboldus2008}
R.~Dotsch and D.~H. Wigboldus, ``Virtual prejudice,'' \emph{Journal of
  Experimental Social Psychology}, 2008.

\bibitem{GroomEtAl2009}
V.~Groom, J.~N. Bailenson, and C.~Nass, ``The influence of racial embodiment on
  racial bias in immersive virtual environments,'' \emph{Social Influence},
  2009.

\bibitem{MccallEtAl2009}
C.~McCall, J.~Blascovich, A.~Young, and S.~Persky, ``Proxemic behaviors as
  predictors of aggression towards black (but not white) males in an immersive
  virtual environment,'' \emph{Social Influence}, 2009.

\bibitem{EastwickAndGardner2009}
P.~W. Eastwick and W.~L. Gardner, ``Is it a game? evidence for social influence
  in the virtual world,'' \emph{Social Influence}, 2009.

\bibitem{Sabanovic2010}
S.~{\v{S}}abanovi{\'c}, ``Robots in society, society in robots,'' \emph{IJSR},
  2010.

\bibitem{CaslerEtAl2013}
K.~Casler, L.~Bickel, and E.~Hackett, ``Separate but equal? a comparison of
  participants and data gathered via amazon’s mturk, social media, and
  face-to-face behavioral testing,'' \emph{Computers in Human Behavior}, 2013.

\bibitem{Suler2004}
J.~Suler, ``The online disinhibition effect,'' \emph{Cyberpsychology \&
  behavior}, 2004.

\bibitem{FoxAndTang2014}
J.~Fox and W.~Y. Tang, ``Sexism in online video games: The role of conformity
  to masculine norms and social dominance orientation,'' \emph{Computers in
  Human Behavior}, 2014.

\bibitem{Herring1996}
S.~Herring, ``Posting in a different voice: Gender and ethics in
  computer-mediated communication,'' \emph{Philosophical perspectives on
  computer-mediated communication}, 1996.

\bibitem{HudsonBruckman2002}
J.~M. Hudson and A.~Bruckman, ``Disinhibition in a cscl environment,'' in
  \emph{Proc. CSCL}, 2002.

\bibitem{MoradiAndHuang2008}
B.~Moradi and Y.-P. Huang, ``Objectification theory and psychology of women: A
  decade of advances and future directions,'' \emph{Psychology of Women
  Quarterly}, 2008.

\bibitem{GlickAndFiske1996}
P.~Glick and S.~T. Fiske, ``The ambivalent sexism inventory: Differentiating
  hostile and benevolent sexism,'' \emph{Journal of Personality and Social
  Psychology}, 1996.

\bibitem{Ruxton2006}
G.~D. Ruxton, ``The unequal variance t-test is an underused alternative to
  student's t-test and the mann--whitney u test,'' \emph{Behavioral Ecology},
  2006.

\bibitem{WotanisMcmillan2014}
L.~Wotanis and L.~McMillan, ``Performing gender on youtube: How jenna marbles
  negotiates a hostile online environment,'' \emph{Feminist Media Studies},
  2014.

\bibitem{EysselAndLoughnan2013}
F.~Eyssel and S.~Loughnan, ````it don't matter if you're black or white''?'' in
  \emph{Proc. ICSR}, 2013.

\bibitem{BrscicEtAl2015}
D.~Brsci{\'c}, H.~Kidokoro, Y.~Suehiro, and T.~Kanda, ``Escaping from
  children's abuse of social robots,'' in \emph{Proc. HRI}, 2015.

\bibitem{SalviniEtAl2010}
P.~Salvini, G.~Ciaravella, W.~Yu, G.~Ferri, A.~Manzi, B.~Mazzolai, C.~Laschi,
  S.-R. Oh, and P.~Dario, ``How safe are service robots in urban environments?
  bullying a robot,'' in \emph{Proc. RO-MAN}, 2010.

\bibitem{BartneckEtAl2005}
C.~Bartneck, C.~Rosalia, R.~Menges, and I.~Deckers, ``Robot abuse -- a
  limitation of the media equation,'' in \emph{Proc. Interact 2005: Workshop on
  Agent Abuse}, 2005.

\bibitem{RosenthalEtAl2013}
A.~M. Rosenthal-von~der P{\"u}tten, N.~C. Kr{\"a}mer, L.~Hoffmann, S.~Sobieraj,
  and S.~C. Eimler, ``An experimental study on emotional reactions towards a
  robot,'' \emph{IJSR}, 2013.

\bibitem{Breazeal2004}
C.~Breazeal, ``Social interactions in hri: the robot view,'' \emph{IEEE
  Transactions on Systems, Man, and Cybernetics}, 2004.

\bibitem{TanEtAl2018}
X.~Z. Tan, M.~V{\'a}zquez, E.~J. Carter, C.~G. Morales, and A.~Steinfeld,
  ``Inducing bystander interventions during robot abuse with social
  mechanisms,'' in \emph{Proc. HRI}, 2018.

\bibitem{JungEtAl2015}
M.~F. Jung, N.~Martelaro, and P.~J. Hinds, ``Using robots to moderate team
  conflict: the case of repairing violations,'' in \emph{Proc. HRI}, 2015.

\end{thebibliography}

\end{document}